%% file: conference_101719.tex
\documentclass[conference]{IEEEtran}
\IEEEoverridecommandlockouts
\usepackage{cite}
\usepackage{amsmath,amssymb,amsfonts}
\usepackage{algorithmic}
\usepackage{graphicx}
\usepackage{textcomp}
\usepackage{xcolor}
\def\BibTeX{{\rm B\kern-.05em{\sc i\kern-.025em b}\kern-.08em
    T\kern-.1667em\lower.7ex\hbox{E}\kern-.125emX}}

\usepackage[hidelinks]{hyperref}
\usepackage{url}

\usepackage{caption}
\usepackage{pifont}
\usepackage{fancybox}
\usepackage{framed}
\usepackage{graphicx}
\usepackage{afterpage}

\newcommand{\xd}[1]{\textcolor{black}{#1}}

\newcounter{GapCounter}
\newcommand{\GapObservation}{
    \stepcounter{GapCounter} 
    \textbf{Gap Observation-\theGapCounter:} 
}

\newcounter{DirectionCounter}
\newcommand{\DirectionProposal}{
    \stepcounter{DirectionCounter} 
    \textbf{Future Direction Proposal-\theDirectionCounter:} 
}

\makeatletter
\newcommand{\linebreakand}{%
  \end{@IEEEauthorhalign}
  \hfill\mbox{}\par
  \mbox{}\hfill\begin{@IEEEauthorhalign}
}
\makeatother

\begin{document}
\pagestyle{plain}
\title{Bridging the Gap: A Study of AI-based Vulnerability Management between Industry and Academia
\thanks{*Xinda Wang is the corresponding author.}
}

\author{\IEEEauthorblockN{Shengye Wan}
\IEEEauthorblockA{Meta Platforms, Inc.\\
simonwan@meta.com}
\and
\IEEEauthorblockN{Joshua Saxe}
\IEEEauthorblockA{Meta Platforms, Inc.\\
joshuasaxe@meta.com}
\and
\IEEEauthorblockN{Craig Gomes}
\IEEEauthorblockA{Meta Platforms, Inc.\\
craiggomes@meta.com}
\and
\IEEEauthorblockN{Sahana Chennabasappa}
\IEEEauthorblockA{Meta Platforms, Inc.\\
csahana@meta.com}\\
\linebreakand
\IEEEauthorblockN{Avilash Rath}
\IEEEauthorblockA{The University of Texas at Dallas\\
avilash.rath@utdallas.edu}
\and
\IEEEauthorblockN{Kun Sun}
\IEEEauthorblockA{George Mason University\\
ksun3@gmu.edu}
\and
\IEEEauthorblockN{Xinda Wang*}
\IEEEauthorblockA{The University of Texas at Dallas\\
xinda.wang@utdallas.edu}
}

\maketitle

\input{0_abstract}
\input{1_intro}

\input{2_industry_opportunity}
\input{3_industry_asks}
\input{4_industry_contributions}
\input{8_conclusion}

\bibliographystyle{plain}
\bibliography{ref.bib}

\end{document}

%% file: 0_abstract.tex
\begin{abstract}
Recent research advances in Artificial Intelligence (AI) have yielded promising results for automated software vulnerability management. AI-based models are reported to greatly outperform traditional static analysis tools, indicating a substantial workload relief for security engineers. However, the industry remains very cautious and selective about integrating AI-based techniques into their security vulnerability management workflow. To understand the reasons, we conducted a discussion-based study, anchored in the authors' extensive industrial experience and keen observations, to uncover the gap between research and practice in this field. We \xd{empirically} identified three main barriers preventing the industry from adopting academic models, namely, complicated requirements of scalability and prioritization, limited customization flexibility, and unclear financial implications. Meanwhile, research works are significantly impacted by the lack of extensive real-world security data and expertise. We proposed a set of future directions to help better understand industry expectations, improve the practical usability of AI-based security vulnerability research, and drive a synergistic relationship between industry and academia.
\end{abstract}






%% file: 1_intro.tex
\section{Introduction}
Artificial Intelligence (AI) for software security has garnered increasing interests from both industry professionals~\cite{IBM, deepmind,darpa_challenge} and academic researchers~\cite{steenhoek2023empirical, thongtanunam2022autotransform, xia2023automated}, establishing itself as a vibrant domain of both realms. Although there are numerous initiatives and projects that offer unique potential, a detailed examination of real-world practices indicates a divergence~\cite{mazuera2021shallow,chakraborty2021deep,steenhoek2023empirical,nong2022open}. Specifically, the direct implementation or adaptation of research-derived AI models in industry security work-streams remains relatively rare.
To build a symbiotic relationship between industry practices and academic theories, we carefully review the industry's main needs in managing vulnerabilities and the common research approaches. 

In Section~\ref{sec:industry}, our paper first reviews the prevailing mechanisms of the industry for managing security vulnerabilities. This exploration is twofold. We summarize the established non-AI methods, which have long served as the foundation of industry workflows. Then, we shift our focus to a review of the representative emergent AI-based methods 
that have recently been proposed.

Delving deeper, we identify three barriers in Section~\ref{sec:barries_industry} \xd{based on our hypothesis}, which prevent the industry from embracing and incorporating academic AI models into their security operations.
Specifically, the three barriers include a noticeable lack of confidence in models generated by academia, reservations about the adaptability of these academic models, and concerns about the financial costs of adopting academic models within current operational systems. We suggest three research directions for future academic AI security researchers to further address and alleviate these concerns. 

Furthermore, we discuss two potential channels through which the industry can significantly contribute to academic endeavors in Section~\ref{sec:barries_academy}. We list two barriers that may prevent such collaborations. Specifically, we emphasize the criticality of accessing real-world industry datasets within academia, and the need for academia to better utilize industry security expertise. Similar to the previous section, we suggest collaboration pathways for both industry and academia to jointly explore, aiming to bridge these divides and achieve mutual benefits.

\begin{figure}[]
\centering
\captionsetup{justification=centering}
\includegraphics[width=1.0\linewidth]
{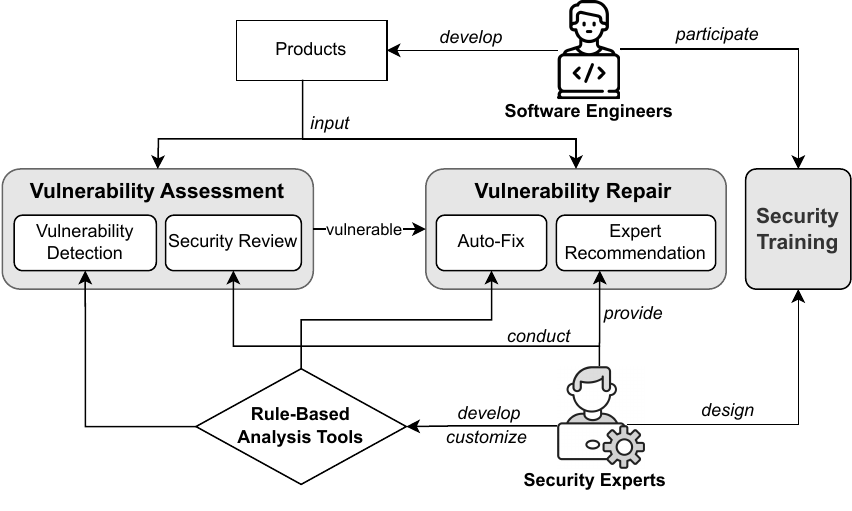}
\caption{\xd{Traditional Industry Workflow of Security Vulnerability Management}}\label{fig:traditional_flow}
    \centering
\end{figure}

In summary, this experience paper makes the following contributions:

\begin{itemize}

\item We outline five gaps that \xd{we believe} currently impede the industry-academia interface in AI-driven security, {with three of these gaps presenting challenges for the industry's adoption of academic advancement, and the other two emphasize challenges in how the industry can beneficially inform and support academia.}

\item We highlight on special challenges and opportunities introduced by AI-based vulnerability management, focusing on key needs like flexible solutions and understanding AI’s cost that are unique for AI-based methods compared with traditional rule-based techniques.

\item Through detailed discussion and structured recommendations, we not only spotlight these areas of divergence but also offer insights and directions for future initiatives, aiming at fostering a more collaborative relationship between industry practices and academic research in the dynamic field of AI-enhanced security.

\end{itemize}

%% file: 2_industry_opportunity.tex
\section{Industry Security Vulnerability Management}
\label{sec:industry}


{This section firstly introduces traditional workflows in industry security risk management, focusing on non-AI methods that have been the cornerstone of protecting digital assets. It then explores how AI technologies can enhance these established processes, offering new ways to strengthen each phase of  workflow against evolving landscape of cyber threats.}


\subsection{Traditional Workflow}
\label{subsec:industry_workflow}
Two main stages in industry security vulnerability management life cycle are \emph{Vulnerability Assessment} and \emph{Vulnerability Repair}, each of which contain multiple steps with different analysis tools. Also, industries offer various kinds of periodic \emph{Security Training} to enhance the process. The entire workflow is depicted in Figure~\ref{fig:traditional_flow}\footnote{This figure selectively presents components directly related to the gaps we discuss in later sections, concentrating our experience paper's content on these specific aspects. \xd{Other components and tools in security vulnerability management, such as feedback loop, vulnerability prioritization, and fuzzing techniques, are omitted for simplicity.}}.

\noindent \textbf{Vulnerability Assessment}. The vulnerability management life-cycle begins with the vulnerability assessment phase. This typically involves a combination of automated \emph{Vulnerability Detection} and manual \emph{Security Review}. 
Automated assessments mainly leverage \emph{rule-based analysis tools}, including dynamic and static analysis, geared towards identifying known vulnerabilities. These tools are typically integrated into the development cycle, continually scrutinizing products throughout their development stages. 

Threat modeling and reach-ability analysis is conducted to accurately evaluate product surface priorities. Surfaces with zero-click or one-click surfaces trigger an additional phase of manual security review process. These manual reviews involve a comprehensive examination of code components, allowing for the precise identification of any hidden risks or vulnerabilities that automated tools may have overlooked. This human-driven security review by security experts adds an essential layer of scrutiny to enhance overall security posture. Security experts, armed with first-hand knowledge, play a crucial role in triaging and unraveling the underlying security issues, providing nuanced insights beyond superficial vulnerabilities.

\noindent \textbf{Vulnerability Repair}. Upon detecting vulnerabilities, security experts utilize industry best practices for remediation. This includes a blend of strategic auto-patch management (\emph{Auto-Fix}), and \emph{Expert Recommendation} through manual code reviews. {We have observed that the mainstream auto-fix solutions are still developed based on different security rules, particularly for common vulnerabilities that can benefit from a unified approach, like a framework, to prevent breaches. Meanwhile, for vulnerabilities that are newly discovered or lack a universal solution, security experts become more involved to offer expert recommendations.}

{It's noted that there is variance in remediation cultures across different teams, reflecting distinct strategic approaches and priorities in addressing and repairing vulnerabilities. In some instances, automated tools are leveraged to speed up the patching of vulnerabilities. Additionally, we've observed a new opportunity for real-time repair solutions offered by Large Language Models. However, concerns about efficiency arise, especially when considering the balance between the accuracy of AI-based suggestions and the engineering effort required to verify these recommendations. More details on this will be discussed below.}

\noindent \textbf{Security Training}. Industrial organizations typically conduct periodic security training events as a strategy to enhance security. These sessions serve a dual purpose of disseminating best practices and providing crucial updates on emerging threats. They equip the product teams with the knowledge and skills necessary to proactively safeguard the organization’s digital assets.

\subsection{AI Opportunities}
In traditional industry security vulnerability management, each stage significantly relies on human security experts, providing opportunities for AI to relieve human efforts as outlined below.

\noindent\textbf{AI for Vulnerability Assessment}. 
Recent deep learning (DL) based methods have shown superior performance in vulnerability assessment tasks such as vulnerability detection, code clone detection, vulnerability severity evaluation. The adopted DL architectures include recurrent neural network (RNN)~\cite{wang2021patchrnn}, graph neural network (GNN)~\cite{mirsky2023vulchecker}, transformer~\cite{fu2022linevul}, and large language model (LLM)~\cite{chen2023diversevul}. These state-of-the-art DL-based approaches have reported more than 90\% F1 scores~\cite{steenhoek2023empirical}, which greatly outperform the traditional static analyzers in most vulnerability assessment tasks. 

Security review, while effective for quality assurance, is laborious and prone to error when security analysts manually assess flaws and offer detailed feedback. Recent research introduced AI-based methods to automate code review comments, supporting the modern code review process~\cite{tufano2022using, thongtanunam2022autotransform}. These results highlight the potential of AI in industrial vulnerability assessment.

Furthermore, constructing security Supervised Fine-Tuning (SFT) datasets also presents an appealing opportunity~\cite{roziere2023code}. Once equipped with ideal security SFT datasets, AI-based tools could gain a more comprehensive understanding of security breaches, threats, and patterns, thereby enabling AI to identify vulnerabilities with greater precision and speed. Additionally, SFT datasets can be utilized in various scenarios, such as significantly enhancing the prediction quality of LLM models.\footnote{Although improving LLM quality is beyond the scope of this paper, highlighting the SFT opportunity adds extra value and impact.}

\noindent \textbf{AI for Vulnerability Repair}. 
AI-based automated program repair (APR) can help developers automatically fix software bugs including security vulnerabilities. Recent large pre-trained language models, trained using billion of text and code tokens, have shown great capability in APR tasks and can even provide patch rankings and correctness checking to suggest more natural fixes~\cite{xia2023automated}.
Also, when repairing a program, developers are usually tired of writing detailed comments and documentation. AI-based code change summarization, that represents code edit operations as graphs, has achieved preliminary progress in generating explanations to help developers understand code changes including security fixes~\cite{dong2022fira}.

{
Additionally, the development of Specialized Language Models (SLMs) tailored for vulnerability repair catches our attention~\cite{Copilot}. Beyond the general benefits that SLMs offer over LLMs, such as enhanced specificity and efficiency, these SLMs can be uniquely designed to incorporate the distinct coding practices and security standards of individual industrial products. They can adapt to preferred repair solutions, effectively aligning with the complicated requirements of different company-internal systems. This capability ensures that suggested fixes are not only technically accurate but also practically applicable, adhering to the unique operational and security protocols of each product.
}


\noindent \textbf{AI for Security Training}. 
The key reason of introducing security flaws during software development is a lack of awareness~\cite{pearce2023examining}. In AI-based security training, each developer is provided with essential knowledge of various types of security flaws according to the level of security they can implement~\cite{li2021vulnerability}. It has demonstrated effectiveness to prevent security problems (e.g., phishing attacks) in established tech companies~\cite{ansari2022prevention}.

Despite significant investments from big industry companies into developing AI-based vulnerability management tools~\cite{IBM, deepmind}, the performance of these tools hasn’t reached the level of computer vision (CV) and natural language processing (NLP) products. Consequently, we haven't observed widespread usage of these tools in industry security vulnerability management workflows. We will explain why adopting this type of research directly in industry is non-trivial in the following sections.

%% file: 3_industry_asks.tex
\section{Barriers to Industry Adoption of Academic Research}
\label{sec:barries_industry}

This section identifies three main observed barriers between academic research and its industry applications, {particularly in the wake of the AI research surge in vulnerability management that has intrigued the industry.} It then proposes corresponding future research directions to bridge these gaps, {aiming to facilitate the integration of the latest research findings into practical applications.}

\subsection{Inconsistent Scope and Priority}
\cornersize{.15}
\noindent{\ovalbox{\begin{minipage}{8.6cm}
\GapObservation Industry tends to favor models that are extremely proficient at addressing certain vulnerabilities (i.e., more severe or exploitable ones), even if they constitute only a subset of all vulnerabilities. In contrast, most research endeavors propose one-for-all solutions with variable performance against different types of vulnerabilities.
\end{minipage}}}
\vspace{0.01in}

\indent \emph{Model Disagreement.} Finding a solution universally superior for addressing all types or most types of vulnerabilities is highly challenging and perhaps unrealistic shortly, as illustrated by recent research by Steenhoek et al.~\cite{steenhoek2023empirical}. Their detailed evaluation shows that leading AI models only agree 7\% of the time across various test data. Even among the top three models, the agreement is less than 50\%. The absence of a unified agreement hinders the adoption of a single, standardized tool for managing vulnerabilities across various organizations.
Moreover, it is common that research works~\cite{croft2023data,chakraborty2021deep} present evaluations using a universal dataset, typically illustrating a rise in total detected issues as performance improvements without delving into the specific areas where the model excels.

\emph{Misaligned Priority.} In mature industries, security teams typically handle various types of vulnerabilities through different project plans or even teams, making it financially and operationally acceptable to utilize different tools for managing vulnerabilities, {where each tool selected is often specialized to address certain vulnerabilities more effectively than others. This diversity in tool usage underscores the industry's practical approach to security, which prioritizes specialized efficacy over a one-size-fits-all solution.}
Under such settings, the fundamental metric for each project and tool is to mitigate a specific type of vulnerability as thoroughly as possible. This crucial context for the practical applications of these tools in the industry is overlooked or insufficiently considered in recent academic research~\cite{chen2023diversevul,fu2022linevul,mirsky2023vulchecker}.
For example, the majority of research efforts only focus on addressing the CWE top 25 most dangerous software weaknesses~\cite{CWEtop25} that are determined through statistical analysis of public vulnerability data sourced from the National Vulnerability Database (NVD), which may be far from the vulnerability distribution {or fixing priority} within specific industrial internal projects. 

\noindent{\ding{226}\DirectionProposal \textbf{\emph{Emphasizing Specialized Model Research.}}}
To increase industry confidence in academic models, there is merit in researchers redirecting their focus towards developing models that specialize in certain types of vulnerabilities, such as buffer overflows. {In a recent empirical study~\cite{steenhoek2023empirical}, the authors experimented and verified that five different types of vulnerabilities achieved the best F1 score across five different models.} This narrower focus aligns with the long history of security research that concentrates on specific issues, such as the automatic detection of memory exploitation~\cite{memory:brumley2008automatic,memory:heelan2019gollum} and SQL injection~\cite{sqli:al2023sqirl,sqli:martin2008automatic}. 
Even for the one-for-all solutions, we hope academia researchers can extensively exercise their models for more vulnerability distributions instead of strictly adhering to the priorities in the CWE Top 25.
Such targeted research would allow for comprehensive evaluations using diverse datasets, thereby offering persuasive evidence of a model’s effectiveness and reliability for specific problems in real-world settings.

\subsection{Limited Customization Flexibility}

\cornersize{.15}
\noindent{\ovalbox{\begin{minipage}{8.6cm}
\GapObservation Industries often need the capability to customize vulnerability management tools for two primary reasons: (i) accommodating varied products and (ii) adhering to different security standards across teams. However, current academic models rarely address these critical customization requirements.
\end{minipage}}}
\vspace{0.01in}

\emph{Public Accessibility of Research Models.} Unlike the CV and NLP domains where models, datasets, and leaderboards from research works are usually open-sourced on GitHub and Hugging Face, most DL-based vulnerability management approaches are not publicly accessible. A recent study~\cite{nong2022open} shows that only a small portion (25.5\%) of the 55 examined papers on DL-based vulnerability detection provided public available tools. 54.5\% available tools contain incomplete documentation and 27.3\% of them have non-functional implementation, making it challenging for industrial users to execute.

\emph{Customization for Diverse Codebases.} 
Ideal industry models should allow for easy customization to suit different products or teams, while maintaining confidence in their predictive outcomes. 
In Section~\ref{subsec:industry_workflow}, we present that within a single corporation, various codebases present unique requirements due to different coding styles, risk tolerances, and the use of distinct compilation tools and configurations across teams.
Since models that are fine-tuned for specific scenarios significantly outperform their generic, pre-trained counterparts in detecting vulnerabilities~\cite{steenhoek2023empirical}, research models should be able to easily fit into above differences in order to operate reliably and accurately across diverse organizational scenarios.

\emph{Integration with and Learning from Existing Tool Ecosystems.}
The industry also stands to benefit from models that are modularized to interface with other security tools effectively. As explained in Section~\ref{subsec:industry_workflow}, conventional industry workflows employ an array of vulnerability management tools, each designed to analyze and assess different aspects of exploitability and the potential impact of identified vulnerabilities. In this environment, it is crucial for newly adopted academic models to integrate seamlessly with existing tools and to possess learning capabilities. This functionality would enable them to iteratively incorporate and apply knowledge derived from an organization's existing suite of security tools.

\vspace{0.03in}
\noindent \ding{226}\DirectionProposal \textbf{\emph{Developing Flexible and Scalable Models.} }
Future research on DL-based vulnerability detection should follow open science practice to provide well-maintained documentation, executable source code, and sufficient design/implementation details for reproducing.
Besides, future research should aim to create models that are flexible enough to be tailored to various products and scalable to adapt to different security standards across teams. Such an approach would facilitate continuous improvement and integration of industry-specific insights, ensuring that the models remain relevant and effective in diverse security environments.

\subsection{Unclear Financial Implications}
\cornersize{.15}
\noindent{\ovalbox{\begin{minipage}{8.6cm}
\GapObservation {While understanding the financial benefits of incorporating AI-based security vulnerability management is crucial for the industry, previous research works inadequately discuss them, even less so in the context of real-world industry scenarios, such as computation power versus human resources.}
\end{minipage}}}
\vspace{0.01in}

\indent \emph{Model Evaluation.}
Academic works~\cite{he2022distribution, croft2023data} focus on the efficacy of models in identifying vulnerabilities, often adopting datasets with unrealistic proportions of vulnerable and non-vulnerable samples (e.g., 50-50~\cite{zhou2019devign}) and neglecting the real-world scenario where benign code significantly outnumbers vulnerable code. When a model identifies a moderate-risk vulnerability but generates ten times as many false positives, it is unlikely to be adopted due to the substantial operational burdens that ultimately undermine its financial viability. Also, with little consideration on explainability, how the models make the correct decisions remains unclear: if the model is able to capture the vulnerable code semantics or just wrongly regard the non-vulnerable part in the vulnerable sample as the vulnerability. The models for latter case are not reliable and will cause financial loss in practice. In summary, the evaluation metrics and scenarios employed in academia provide limited insight into financial impacts of integrating these models into organizational systems.

\emph{Scalability and Variability Concerns.}
The uncertainty surrounding the financial implications of adopting academic models is further compounded by their untested performance on extensive and diverse industry codebases and infrastructures. 
For instance, an existing study~\cite{steenhoek2023empirical} has observed more than 50\% performance drop when applying academic models to more complex open-source software datasets.
Language models like Claude-2 and GPT-4 can only solve 4.8\% and 1.7\% real-world GitHub issues, respectively~\cite{jimenez2023swe}.
This variability raises concerns about cost-effectiveness of deploying such models on even more complex proprietary codebases in industry.
On the other hand, recent popular large models with billion-level parameter size limit the development of research works conducted in academic labs with restricted GPU resources. The potentials and capability of large models for vulnerability management are still awaiting for systematic and scientific exercise and evaluation from academic perspective.

\vspace{0.03in}
\noindent \ding{226}\DirectionProposal \textbf{\emph{Constructing Industry Reflective Evaluation Metrics, Datasets, and Resources for Effective Assessment}.} 
Upcoming academic investigations should aim to develop evaluation metrics and scenarios that align more closely with authentic industry environments, such as incorporating a wider array of variables. This initiative, requiring efforts from both industry and academia, will not only evaluate models' effectiveness in a more fitting industry context but also showcase their practical and cost-effective use in various real-world scenarios. Specifically, one industry-context evaluation could involve integrating the savings in salary or labor costs into the analysis of models' performances.


%% file: 4_industry_contributions.tex
\section{Barriers to Academia Adaptation of Industry Settings}
\label{sec:barries_academy}
This section aims to explore the potential contributions that the industry can make to academia, as well as the barriers that need to be overcome to facilitate this exchange. {Specifically, it looks into how industry insights and resources can help academic researchers enhance the real-world applicability and performance of their work.}

\subsection{Shortage of Large-Scale and Diverse Datasets}
\cornersize{.15}
\noindent{\ovalbox{\begin{minipage}{8.6cm}
\GapObservation The extensive databases within industries hold invaluable insights into insecure practices across various stages of development. However, the sharing of these datasets is significantly hindered by many organizations' risk of unintentional disclosure of sensitive information.
\end{minipage}}}
\vspace{0.01in}

\indent Prior research has shown that ML-based vulnerability mitigation solutions can achieve better performance from larger and more diverse training dataset~\cite{steenhoek2023empirical, mirsky2023vulchecker} from the following perspectives.

\textit{More Vulnerability Samples.} 
Synthetic datasets, widely used in learning-based vulnerability management tasks to address the limitation of real world data, are usually generated by keeping vulnerable code unchanged and adding variations to unrelated neighboring code.
Since program slicing techniques are used to identify lines of code considered most relevant to potential software vulnerabilities, synthetic datasets introduce huge duplicate slices~\cite{allamanis2019adverse}.
The model trained with such unrealistic synthetic datasets lead to more than 50\% performance drop in practice~\cite{chakraborty2021deep}. To prevent this, one solution is to include real-world datasets with vulnerability complexity and varieties as much as possible.



\textit{More Accurate Labeling.}
With high standards, industry vulnerability management process preserves the quality of vulnerability data. In contrast, up to 70\% vulnerability labels in open-source GitHub repositories are inaccurate~\cite{chakraborty2021deep, wang2019detecting}.
This is because, while some public GitHub repositories permit direct commits with minimal review, industry practices, particularly within large tech corporations, incorporate multiple layers of coding quality safeguards. These safeguards include mandatory peer reviews prior to committing, periodic codebase optimization initiatives led by technical leaders, and automated improvements executed by quality and security bots. 
Learning-based vulnerability mitigation systems can greatly benefit from these high-quality data in industry datasets.

\textit{Wider View of Vulnerabilities.}
In the industry, the traces of detecting each vulnerability are well documented, including security analyst's strategies and general business logic.
They provide more context of detecting decisions to enable a wider view to examine the vulnerabilities than simple vulnerable-secure code pairs in datasets presented by research works.

\textit{Multimodal Information.} Industry datasets often include different levels of code (e.g., source code, binary code, IR code) and diverse types of documentation (e.g., code comments, reviews, discussions). These multimodal information enables DL-based systems to better understand the semantics of potentially vulnerable code.

\vspace{0.03in}
\noindent{\ding{226}\DirectionProposal{ \textbf{\textit{Exploring Data Anonymization and Environment Simulation Techniques.}}} 
\xd{Despite the advantages of sharing datasets, it is worth noting that potential negative effects may arise from sharing raw industry data. Attackers may leverage identifying information from these newly disclosed data to architect novel attacks. Thus, developing advanced methods for effective data anonymization is essential to guarantee the secure sharing of industry internal datasets, particularly code segments.} 
Industries, willing to contribute historical security events and legacy vulnerable code to academia, seek assurance that sharing codebases will not expose sensitive and identifying information, necessitating thorough anonymization of shared data. Academia, on the other hand, requires data that retains as much of the original patterns and contexts of vulnerabilities after anonymization. This will enable researchers to explore secure mechanisms for sharing vulnerability contexts further.
Additionally, in situations where sharing anonymized data is still considered risky, another option is allowing researchers access to industry settings. \xd{In case of bad actors or potential misuse,} these settings could either mirror or emulate the actual processes associated with vulnerability management in an industrial framework \xd{with necessary access control and data isolation}.

\subsection{Lack of First-Hand Industry Expertise}
\cornersize{.15}
\noindent{\ovalbox{\begin{minipage}{8.6cm}
\GapObservation Access to first-hand industry expertise is pivotal for enhancing effectiveness of research models in practice; however, obtaining such expert knowledge and experience remains a significant challenge for most academic researchers.
\end{minipage}}}
\vspace{0.01in}

\emph{Advantages of practical expertise in AI-based system design.} While deep learning effectively automates feature extraction in traditional CV and NLP tasks, security-related tasks often necessitate expert involvement in crafting appropriate data representations and AI algorithms~\cite{alon2019code2vec}. For instance, for optimal feature extraction, researchers transform program source code into various forms, such as code sequence~\cite{fu2022linevul}, natural code sequence graphs~\cite{mirsky2023vulchecker}, patch code property graphs~\cite{wang2023graphspd}, or function behavior graphs~\cite{yuan2023enhancing}, before input into deep learning models. However, the creation of these representations may not fully encapsulate the complexities of industrial programming in large proprietary projects due to researchers typically relying on specific popular open-source projects (e.g., Linux kernel, FFmpeg, and QEMU) as references~\cite{wang2021patchdb}. This reliance may result in a limited understanding of intricate, real-world security issues. 
Since cutting-edge projects often begin as proprietary before possibly transitioning to open-source, insights derived from existing open-source projects might not be timely or relevant to the constantly evolving landscape of security threats. Actively engaging industry practitioners in  developing AI-based security systems could bridge this knowledge gap. 


\vspace{0.03in}
\noindent \ding{226}\DirectionProposal \textbf{\emph{Fostering Bidirectional Collaboration between Industry and Academia.}} 
To effectively harness industry expertise in academia, maintaining a robust knowledge exchange channel is crucial.
Industries could support programs that offer internship opportunities and invite more esteemed academic researchers for joint research initiatives.
This collaboration enables academics to access significant resources and lets industry direct research towards practical solutions meeting their needs.

Industries may also consider developing open-source comprehensive testing benchmarks to embed industry's unique requirements for research works, such as the recent release of the Purple LLama CyberSecEval benchmark~\cite{cyberseceval}. Currently, 
there's a lack of a security-focused representative benchmark that comprehensively and accurately reflects the performance of all models. 
As an ideal outcome, we could expect a robust, industry-wide benchmark dataset, similar to the impactful precedent set by ImageNet~\cite{deng2009imagenet}, which would enable the comparison of competing approaches across global academic research streams.
Moreover, industries can host more technical conference (e.g., GitHub Universe\cite{githubuniverse}) from the perspective of developers to bring together both professionals and researchers for  better learning experiences from each other. 

It would be better for security research conferences to encourage the participation of industry developers by calling for practical experience papers, setting up industry tracks, and hosting workshops and tutorial sessions. By gathering invaluable inputs from industry, researchers will be able to improve their work for practical use.



%% file: 8_conclusion.tex
\section{Discussion}

There has been a line of survey papers~\cite{lin2020software,hanif2021rise} on DL-based vulnerability management. The primary goal of this paper is not to provide a comprehensive review on existing works. 
Since this work aims to highlight new challenges in AI-based methods, we only talk about the representative existing works instead of thoroughly listing the literature.
Moreover, we summarize five gaps between the research and practice and propose a set of future direction, which are rarely covered by previous works.

This paper is primarily based on the perspective of industrial practitioners, articulating their concerns and expectations for future research works on AI-based vulnerability management. 
The main authors of this paper are security engineers/researchers from established tech companies, their analysis results may embody biases, as different industrial companies might emphasize specific security aspects based on their business scope and objectives.
However, we believe our findings and proposed directions address prevalent problems shared within the industry, applying to most modern security vulnerability management practices~\cite{practice}. 
We hope this experience paper can offer fresh industrial viewpoints and aid academic research in grasping practical industry demands more effectively.

The five \xd{hypotheses} presented in this paper are encapsulated from the immediate experiences of industrial practitioners, \xd{based on direct half-hour one-on-one interviews with ten practitioners, and supplemented by team meetings and discussion materials kindly shared by these practitioners. The practitioners are supported by four teams, each comprising around ten security experts, specializing in the domains listed in Figure~\ref{fig:traditional_flow}, such as vulnerability detection and auto-fix. 
} 
Also, we focus on several typical vulnerability management approaches as listed in Figure~\ref{fig:traditional_flow}. We leave the discussion on additional components and tools of security vulnerability management (e.g., feedback loop, vulnerability prioritization, fuzzing techniques, etc.) as our future work. 

\xd{There may be potential downsides of our proposed future directions. For example, bad actors may leverage AI to attack vulnerabilities newly disclosed in industry datasets. But we believe that, with the powerful AI-based vulnerability mitigation tools developed from large-scale industry data, defenders can take one step ahead to eliminate their vulnerable code, therefore cutting off the possibility of being attacked. 
Also, research works that offer wider scope, greater customization flexibility, and financial analysis within industry scenarios may require extra workload and not receive adequate recognition during the paper review process. We anticipate mechanisms such as artifact review and badging will encourage academic researchers to focus more on these practical aspects. }

\section{Conclusion}
In this paper, we summarize our direct experiences in integrating AI-based vulnerability management research work to real-world industrial environments.
We empirically identify five distinct gaps between the industry and academia, which include three limitations of the current research that limit the usability of AI-based techniques and two challenges that the industry faces in contributing more effectively to academic inquiries.
To mitigate these gaps, we propose five future directions for both academia and industry to collaboratively forge more effective and practical AI-based vulnerability management tools as well as better leverage informative industry datasets and expertise.
